\documentclass[prc,aps,twocolumn]{revtex4}
\usepackage{graphicx} 
\begin{document}

\title{Quark Structure of the Nucleon and 
Angular Asymmetry of  Proton-Neutron Hard Elastic 
Scattering}

\author{Carlos~G. Granados and Misak~M.~Sargsian}

\affiliation{Florida International University, Miami, FL 33199 USA}

\date{\today}

\begin{abstract}
We investigate an asymmetry in the angular distribution 
of hard elastic proton-neutron scattering with respect to 90$^0$ center of 
mass scattering angle. 
We demonstrate that the magnitude of the angular asymmetry is related 
to  the helicity-isospin symmetry of the quark wave function of the nucleon.
Our estimate of the asymmetry within the quark-interchange model
of hard scattering demonstrates that the quark wave function of a nucleon 
based on the exact SU(6) symmetry predicts an angular asymmetry opposite 
to that of  experimental observations. On the other hand the quark wave 
function based on the diquark picture of  the  nucleon produces an 
asymmetry  consistent with the data. Comparison with the data allowed us 
to extract the relative sign and the magnitude of the vector and scalar 
diquark components of the  quark wave function of the nucleon. These 
two quantities are essential in constraining QCD models of a nucleon.
Overall, our conclusion is that the angular asymmetry of a hard elastic  
scattering of baryons provides a new venue in probing 
quark-gluon structure of baryons and should be considered as an 
important  observable in constraining the theoretical models.
\end{abstract}

\maketitle

For several decades elastic nucleon-nucleon scattering at high momentum transfer 
($-t,-u\ge  M_N^2$~GeV$^2$) has been one  of the important testing grounds for 
QCD dynamics of the strong interaction between hadrons.   Two major observables 
considered were the energy dependence of the elastic cross section 
and the polarization properties of the reaction.

Predictions for energy dependence are based on the underlying dynamics of 
the hard scattering of quark components of the  nucleons. 
One such prediction is based on the quark-counting rule
\cite{BF75,MMT} according to which the differential 
cross section of two-body elastic scattering ($ab\rightarrow cd$) 
at high momentum transfer behaves like 
${d\sigma\over dt} \sim s^{-(n_a + n_b + n_c + n_d)}$, where 
$n_i$ represents the number of constituents in particle $i$ (i=a,b,c,d).

For elastic $NN$ scattering,  the quark-counting rule  predicts 
$s_{NN}^{-10}$ scaling which agrees reasonably well 
with experimental measurements 
(see e.g. Refs.\cite{Akerlof,Allaby,pnexp1,pnexp2}). 
In addition to energy dependence, the comparison~\cite{h20} of the cross 
sections of hard exclusive scattering of hadrons containing quarks  
with the same  flavor  with the scattering of hadrons that share no 
common flavor of quarks demonstrated that the quark-interchange represents the 
dominant mechanism of hard elastic scattering for up to ISR energies 
(see discussion in \cite{BCL}).

For polarization observables, the  major prediction of the QCD dynamics 
of hard elastic scattering 
is the conservation of helicities of  interacting hadrons. The latter 
prediction is based on the fact that the gluon exchange  
in massless quark limit conserves the helicity of interacting quarks.

Quark counting rule and helicity conservation however do not describe 
completely the features of hard scattering data. The energy dependence of 
$pp$ elastic cross section scaled by $s_{NN}^{10}$ exhibits an oscillatory 
behavior which indicates the existence of other possibly nonperturbative mechanisms 
for the scattering\cite{pire,BT}.   These expectations are reinforced also 
by the observed large asymmetry, $A_{nn}$ at some 
hard scattering kinematics\cite{Krabb} which 
indicates  an anomalously large contribution from double helicity flip 
processes.
These  observed discrepancies however do not represent the dominant 
features of the data and overall one can conclude that  the bulk of the 
hard elastic $NN$ scattering amplitude  is defined by the exchange mechanism of  
valence quarks which interact through the hard gluon exchange 
(see e.g. Refs.\cite{FGST,BCL}). Quark-interchange mechanism also reasonably well 
describes the $90$ c.m. hard  break-up  of two nucleons from the deuteron\cite{gdpn,gdpnpol}.

However,  the energy dependence of a hard scattering cross section, 
except for the verification of the   
dominance of  the minimal-Fock component of the quark wave function of 
nucleon,  provides rather limited information about the symmetry properties of 
the valence quark  component of the nucleon wave function.

In this work we demonstrate that an  observable such as the asymmetry of 
a hard elastic proton-neutron scattering with respect to $90^0$ c.m. scattering may 
provide a new insight into the  helicity-flavor symmetry of 
the quark wave function  of the nucleon. Namely we consider
\begin{equation}
A_{90^0}(\theta) = 
{\sigma(\theta) - \sigma(\pi -\theta)\over \sigma(\theta) + \sigma(\pi-\theta)},
\label{Asym}
\end{equation}
where $\sigma(\theta)$ - is the differential cross section of the elastic 
$pn$ scattering. 
We will discuss this asymmetry in the hard kinematic regime 
in which the energy dependence of the cross section is $\sim s^{-10}$.  
Our working assumption is the dominance of the quark-interchange  
mechanism~(QIM) in the $NN$ elastic scattering at these kinematics. 

\begin{figure}[ht]
\centering\includegraphics[scale=0.6]{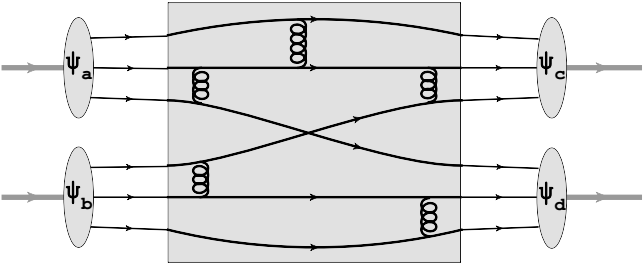}
\vspace{-0.2cm}
\caption{Typical diagram for quark-interchange mechanism of $NN\rightarrow NN$ scattering.}
\label{QIM}
\end{figure}

Within QIM the characteristic scattering diagram can be represented as in Fig.\ref{QIM}.
Here one assumes a factorization of the soft part of the reaction 
in the form of the initial and final state wave functions of nucleons and the hard 
part which is characterized by QIM scattering that proceeds with five hard gluon exchanges 
which generate energy dependence in accordance to the quark counting rule.  
In order to attempt to calculate the absolute cross section of the reaction 
one needs to sum hundreds of diagrams similar to one of Fig.\ref{QIM}. 
However for the purpose of estimation of the asymmetry in Eq.(\ref{Asym}) the 
important observation is that the hard scattering kernel is flavor-blind and 
conserves the helicity. As a result one expects that angular asymmetry will 
be generated mainly through the underlying spin-flavor symmetry of the quark wave functions 
of the  interacting nucleons.

The amplitude of the hard elastic $a+b\rightarrow c+d$ scattering of Fig.\ref{QIM}, 
within quark-interchange approximation, can  be presented as follows:
\begin{widetext}
\begin{eqnarray}
\langle cd\mid T\mid ab\rangle & = &  
\sum\limits_{\alpha,\beta,\gamma} 
\langle  \psi^\dagger_c\mid\alpha_2^\prime,\beta_1^\prime,\gamma_1^\prime\rangle
\langle  \psi^\dagger_d\mid\alpha_1^\prime,\beta_2^\prime,\gamma_2^\prime\rangle 
\nonumber \\
& & \times
\langle \alpha_2^\prime,\beta_2^\prime,\gamma_2^\prime,\alpha_1^\prime\beta_1^\prime
\gamma_1^\prime\mid H\mid
\alpha_1,\beta_1,\gamma_1,\alpha_2\beta_2\gamma_2\rangle\cdot 
\langle\alpha_1,\beta_1,\gamma_1\mid\psi_a\rangle
\langle\alpha_2,\beta_2,\gamma_2\mid\psi_b\rangle,
\label{ampl}
\end{eqnarray}
\end{widetext}
where ($\alpha_i, \alpha_i^\prime$), ($\beta_i,\beta_i^\prime$) and ($\gamma_i,\gamma_i^\prime$) 
describe the spin-flavor quark states before and after the hard 
scattering, $H$,  and 
\begin{equation}
C^{j}_{\alpha,\beta,\gamma} \equiv \langle\alpha,\beta,\gamma\mid\psi_j\rangle
\label{Cs}
\end{equation}
describes the probability amplitude of finding the $\alpha,\beta,\gamma$ helicity-flavor 
combination of three valence quarks in the nucleon $j$\cite{FGST}.

To be able to calculate $C^{j}_{\alpha,\beta,\gamma}$ factors one  
represents the nucleon wave function through the helicity-flavor basis of the valence  
quarks.  We use a rather general form separating  the wave function into two parts characterized 
by two (e.g. second and third)  quarks being  in spin zero - isosinglet and spin one - isotriplet states 
as follows:
\begin{widetext}
\begin{eqnarray}
& & \psi^{i^3_{N},h_N} = {1\over \sqrt{2}}\left\{
\Phi_{0,0}(k_1,k_2,k_3)
(\chi_{0,0}^{(23)}\chi_{{1\over2},h_N}^{(1)})\cdot
(\tau_{0,0}^{(23)}\tau_{{1\over 2},i_N^{3}}^{(1)}) 
\right.  +  \Phi_{1,1}(k_1,k_2,k_3) 
 \times \nonumber \\
& & 
\left.
\sum\limits_{i_{23}^3=-1}^{1} \ \ \sum\limits_{h_{23}^3=-1}^{1}
\langle 1,h_{23}; {1\over 2},h_{N}-h_{23}\mid {1\over 2},h_N\rangle
\langle 1,i^3_{23}; {1\over 2},i^3_{N}-i^3_{23}\mid {1\over 2},i^3_N\rangle
(\chi_{1,h_{23}}^{(23)}\chi_{{1\over2},h_N-h_{23}}^{(1)})\cdot
(\tau_{1,i^3_{23}}^{(23)}\tau_{{1\over 2},i_N^{3}-i^3_{23}}^{(1)})\right\},
\label{wf}
\end{eqnarray}
\end{widetext}
where $j_N^3$ and $h_N$ are the isospin component  and the helicity of the nucleon.
Here  $k_i$'s are the light cone momenta of quarks which should be understood as  
($x_i,k_{i\perp}$) where $x_i$ is a light cone momentum fraction of the nucleon 
carried by the $i$-quark. 
We define $\chi_{j,h}$ and $\tau_{I,i^3}$ as helicity 
and isospin  wave functions, where $j$ is the spin, $h$ is the helicity, 
$I$ is the isospin and $i^3$ its third component.
The Clebsch-Gordan coefficients are 
defined as $\langle j_1,m_1;j_2,m_2\mid j,m\rangle$. 
Here, $\Phi_{I,J}$ represents the momentum dependent part of the wave function 
for ($I=0,J=0$) and ($I=1,J=1$) two-quark spectator states respectively.
Since the asymmetry in Eq.(\ref{Asym}) does not 
depend on the absolute normalization of the cross section,  
a more relevant quantity for us will be
the relative strength of these two
momentum dependent wave functions.  For our discussion we introduce a 
parameter, $\rho$: 
\begin{equation}
\rho = {\langle \Phi_{1,1}\rangle \over \langle \Phi_{0,0}\rangle }
\label{rho}
\end{equation}
which characterizes an average relative  magnitude of 
the wave function components corresponding to 
($I=0,J=0$) and ($I=1,J=1$) quantum  numbers  of two-quark ``spectator'' states.
Note that the two extreme values of $\rho$ define two well know approximations:
$\rho=1$ corresponds to the exact SU(6) symmetric picture of the nucleon 
wave function  and 
$\rho=0$ will correspond to the contribution of only good-scalar diquark configuration 
in the nucleon wave function (see e.g. Ref.\cite{Ansel,RJ,SW,BCR} where this component is referred as 
a scalar or good diquark configuration~($[qq]$) as opposed to a vector or bad diquark 
configuration denoted by  $(qq)$).  
In further discussions we will keep $\rho$ as a free parameter.

To calculate the scattering amplitude of Eq.(\ref{ampl}) we assume a 
conservation of the helicities of quarks participating in the hard scattering. 
This allows us to approximate the hard scattering part of the amplitude, $H$,  
in the following form:
\begin{equation}
H \approx \delta_{\alpha_1\alpha_1^\prime}\delta_{\alpha_2\alpha_2^\prime}
\delta_{\beta_1,\beta_1\prime}
\delta_{\gamma_1,\gamma_1^\prime}
\delta_{\beta_2,\beta_2\prime}
\delta_{\gamma_2,\gamma_2^\prime} {f(\theta)\over s^4}.
\label{H}
\end{equation}
Inserting this expression into Eq.(\ref{ampl}) for the QIM 
amplitude one obtains\cite{FGST}:
\begin{equation}
\langle cd\mid T\mid ab\rangle = Tr(M^{ac}M^{bd})
\label{ampl2}
\end{equation}
with:
\begin{equation}
M^{i,j}_{\alpha,\alpha^\prime} = 
C^{i}_{\alpha,\beta\gamma}C^{j}_{\alpha^\prime,\beta\gamma} + 
C^{i}_{\beta\alpha,\beta}C^{j}_{\beta\alpha^\prime,\beta} + 
C^{i}_{\beta\gamma\alpha}C^{j}_{\beta\gamma\alpha^\prime},
\label{QIMMs}
\end{equation} 
where we sum over the all possible values of $\beta$ and $\gamma$.
Furthermore, we separate the energy dependence from the scattering amplitude as
follows:
\begin{equation}
\langle cd\mid T\mid ab\rangle = {\langle h_c,h_d\mid T(\theta)\mid h_a,h_b\rangle \over s^4}
\label{phidef}
\end{equation}
and define five independent angular parts of the helicity amplitudes as:  
\begin{eqnarray}
& & \phi_1 = \langle ++\mid T(\theta)\mid ++\rangle; \ \ 
 \phi_2 =  \langle --\mid T(\theta)\mid ++\rangle; \nonumber \\ 
& &  \phi_3 =  \langle +-\mid T(\theta)\mid +-\rangle; \ \
\phi_4 =  - \langle -+\mid T(\theta)\mid +-\rangle; \nonumber \\
& & \phi_5 =  \langle -+\mid T(\theta)\mid ++\rangle.
\label{phis}
\end{eqnarray}
Here the ``-'' sign in the definition of $\phi_4$ follows from the 
Jacob-Wick helicity convention\cite{JW} according to which a (-1) phase 
is introduced if two quarks that scatter to $\pi-\theta_{cm}$ angle have
opposite helicity (see also Ref.\cite{FGST}).

Using Eqs.(\ref{ampl2},\ref{QIMMs})  for the non-vanishing helicity 
amplitudes of Eq.(\ref{phis}) one obtains:\\
for $pp\rightarrow pp$:
\begin{eqnarray}
\phi_1 & = &    (3 + y)F(\theta)    +  (3 + y)  F(\pi-\theta) \nonumber \\
\phi_3 & = &    (2 - y)F(\theta)    +  (1 + 2y) F(\pi-\theta) \nonumber  \\
\phi_4 & = &   -(1 + 2y)F(\theta)  -  (2 - y)  F(\pi-\theta)
\label{pppp}
\end{eqnarray}
and for  $pn\rightarrow pn$:
\begin{eqnarray}
\phi_1 & = &  (2 - y)F(\theta)   +  (1 + 2y)  F(\pi-\theta) \nonumber \\
\phi_3 & = &  (2 + y)F(\theta)   +  (1 + 4y) F(\pi-\theta) \nonumber \\
\phi_4 & = &   2y F(\theta)      +     2y  F(\pi-\theta)
\label{pnpn}
\end{eqnarray}
with $\phi_2=\phi_5=0$ due to helicity conservation. Here:
\begin{equation}
y = x(x+1) \ \ \mbox{with } x = {2\rho \over 3(1+\rho^2)} 
\label{xy}
\end{equation}
and $F(\theta)$ is the angular function. Note that the 
$\rho=1$ case reproduces the SU(6) result of Refs.\cite{FGST} and \cite{BCL}.
The results of Eqs.(\ref{pppp}) and (\ref{pnpn}) could be obtained 
also through the formalism of the H-spin introduced in Ref.\cite{BCL}. 
In this case the helicity amplitudes will be expressed through the average number 
of quarks to be found in a given helicity-spin state. These numbers will 
be directly defined through the wave function  of Eq.(\ref{wf}).

Introducing the symmetric and antisymmetric parts of the angular function $F$ as follows:
\begin{equation}
s(\theta) = {F(\theta) + F(\pi-\theta)\over 2}; \ \ 
a(\theta) = {F(\theta) - F(\pi-\theta)\over 2}
\label{as}
\end{equation}
and using  Eq.(\ref{pnpn}) for the asymmetry as it is defined in Eq.(\ref{Asym}) one obtains:
\begin{equation}
A_{90^0}(\theta) = {6 a(\theta)s(\theta)(1-2y - 3y^2)\over 
a(\theta)^2 (1-3y)^2 + 3s(\theta)^2(3 + 6y + 7y^2)}.
\label{asym}
\end{equation}

\begin{figure}[t]
\centering\includegraphics[scale=0.4]{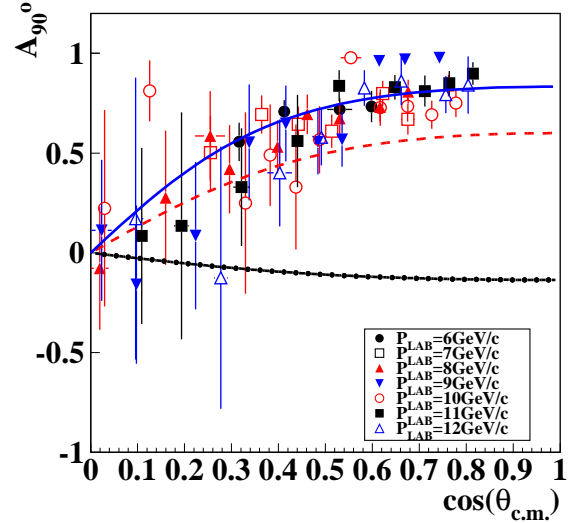}
\vspace{-0.1cm}
\caption{Asymmetry of $pn$ elastic cross section. Solid dotted line - SU(6), with $\rho=1$,
dashed line diquark-model with $\rho=0$, solid line - fit with $\rho=-0.3$.}
\label{Fig.2}
\end{figure}
One can make a rather general observation from  Eq.(\ref{asym}), that for the 
SU(6) model, ($\rho=1$, $y = {4\over 9}$) and for any positive function, $a(\theta)$ at 
$\theta \le {\pi\over 2}$, the  angular asymmetry  has a negative sign opposite 
to the  experimental asymmetry~(Fig.\ref{Fig.2}).  Note that one expects 
a positive $a(\theta)$ at $\theta \le {\pi\over 2}$ from general grounds based on 
the expectation that  in the hard scattering regime the number of $t$-channel 
quark scatterings  dominates the number of $u$-channel quark scatterings  in the forward 
direction.  

As it follows from Eq.(\ref{asym}), positive asymmetry can be achieved 
only for $1-2y-3y^2 >0$, which according to Eq.(\ref{xy}) imposes the following restrictions 
on $\rho$: $\rho<0.49$ or $\rho>2.036$. 
The first condition indicates on the preference of scalar diquark-like configurations in the 
nucleon wave function, while the second one will indicate the strong dominance of the vector-diquark 
component which contradicts the observations\cite{Ansel,RJ,SW}.

In Fig.\ref{Fig.2} the asymmetry 
of $pn$ scattering calculated with SU(6)~($\rho=1$) and pure scalar-diquark~($\rho=0$) models are 
compared with the data.  In these estimates we use 
$F(\theta) = C\cdot sin^{-2}(\theta)(1-cos(\theta))^{-2}$ 
dependence of the angular function\cite{RS} which is consistent with the 
picture of hard collinear QIM scattering of valence quarks with five gluon exchanges and  
reasonably well reproduces the main characteristics  of the  angular dependencies  of 
both $pp$ and $pn$ elastic scatterings. Note that using a form of the angular function 
based on nucleon form-factor arguments\cite{BCL,FGST}, $F\approx (1-cos(\theta))^{-2}$
will result in the same angular asymmetry.

The comparisons show that the nucleon wave function~(\ref{wf}) with a good-scalar 
diquark component~($\rho=0$) produces the right sign for the angular asymmetry.
On the other hand  even large errors of the data do not 
preclude to conclude that the exact SU(6) symmetry~($\rho=1$) of the quark wave 
function of nucleon is in qualitative disagreement with the experimental asymmetry.

Using the above defined angular function $F(\theta)$ we fitted $A_{90^0}$ in  
Eq.(\ref{asym}) to the data at  $-t,-u\ge 2$~GeV$^2$ varying $\rho$ as a free parameter.  
We used the Maximal Likelihood method of fitting excluding those data points from the data set  
whose errors are too large for meaningful identification of the asymmetry. 
The best fit is found for 
\begin{equation}
\rho \approx  -0.3\pm 0.2.
\label{rhofit}
\end{equation}
The nonzero magnitude of $\rho$ indicates  the small but finite relative strength of 
a bad/vector diquark  configuration in the nuclear wave function as compared 
to the scalar diquark component.   
It is intriguing that the obtained magnitude of $\rho$ is consistent with  the $10\%$ 
probability of ``bad'' diquark configuration discussed in Ref.\cite{SW}.

Another interesting property of Eq.(\ref{rhofit}) is the negative sign of the  
parameter $\rho$.

Within  qualitative  quantum-mechanical picture, the 
negative sign of $\rho$  may indicate for example the 
existence of a repulsion in the quark-(vector- diquark) channel as opposed to 
the attraction in the quark - (scalar-diquark) channel.  It is rather surprising that 
both the magnitude and sign agree with the result of the phenomenological interaction 
derived in the one-gluon exchange quark model discussed in Ref.\cite{RJ}.

In conclusion, we demonstrated that the angular asymmetry of hard elastic $pn$ 
scattering can be used to probe the symmetry structure of the valence quark wave 
function of the nucleon. We demonstrated that the exact SU(6) symmetry does not reproduce 
the experimental angular asymmetry of hard elastic $pn$ scattering.
Nucleon wave function consistent with the diquark structure gives a  
right asymmetry. The fit to the data  indicates  $10\%$ probability 
for the existence of bad/vector diquarks in the wave function of nucleons.
It also  shows that the vector  and scalar $qq$ components of the  wave function  may be 
in the opposite phase.  This will indicate on different dynamics of 
$q-[qq]$ and $q-(qq)$ interactions.

The relative magnitude and the sign of the 
vector $(qq)$ and scalar $[qq]$
components can be used to constrain the different QCD  predictions which require the existence 
of diquark components in the nucleon wave function. These quantities in  principle  can be 
checked in Lattice calculations.  The angular asymmetry studies can be extended also to include 
the scattering of other baryons such as  $\Delta$-isobars (which may have a larger fraction of 
vector diquark component) as well  as strange baryons which will allow us to study the relative 
strength of $(qq)$ and $[qq]$ configurations involving strange quarks.

This work is supported by U.S. Department of Energy grant under contract DE-FG02-01ER41172.

\end{document}